\documentclass[preprint]{aastex6}

\usepackage{natbib}
\usepackage{xcolor}
 
\newcommand\species[2]{#1 {\sc #2}}

\def\ie{\mbox{i.e.}}
\def\eg{\mbox{e.g.}}

\def\teff{\mbox{T$_{\rm eff}$}}
\def\logg{\mbox{log~{\it g}}}
\def\vmicro{\mbox{$\xi_{\rm t}$}}
\def\kmsec{\mbox{km~s$^{\rm -1}$}}

\shorttitle{RRc Stellar Spectra}
\shortauthors{Sneden et al.}

\begin{document}

\title{Metal-Rich RRc stars in the Carnegie RR Lyrae Survey}

\author{Christopher Sneden\altaffilmark{1},
        George W.Preston\altaffilmark{2},
        Juna A. Kollmeier\altaffilmark{2},
        Jeffrey D. Crane\altaffilmark{2}, 
        Nidia Morrell\altaffilmark{3}, 
        Jos{\' e} L. Prieto\altaffilmark{4,5},
        Stephen A. Shectman\altaffilmark{2}, 
        Dorota M. Skowron\altaffilmark{6}, 
        Ian B. Thompson\altaffilmark{2}, 
}

\altaffiltext{1}{Department of Astronomy and McDonald Observatory, 
                 The University of Texas, Austin, TX 78712, USA; 
                 chris@verdi.as.utexas.edu}
\altaffiltext{2}{The Observatories of the Carnegie Insitution for Science, 
                 813 Santa Barbara Street, Pasadena, CA 91101, USA; 
                 crane@carnegiescience.edu, gwp@carnegiescience.edu, 
                 jak@carnegiescience.edu, ian@carnegiescience.edu, 
                 nmorrell@carnegiescience.edu, shec@carnegiescience.edu}
\altaffiltext{3}{Las Campanas Observatory, Carnegie Observatories, 
                 Casilla 601, La Serena, Chile; nmorrell@carnegiescience.edu}
\altaffiltext{4}{N\'ucleo de Astronom\'ia de la Facultad de Ingenier\'ia 
                 y Ciencias, Universidad Diego Portales, Av. Ej\'ercito 441, 
                 Santiago, Chile; jose.prietok@mail.udp.cl}
\altaffiltext{5}{Millennium Institute of Astrophysics, Santiago, Chile}
\altaffiltext{6}{Warsaw University Observatory, Aleje Ujazdowskie 4, 
                 00-478 Warszawa, Poland; dszczyg@astrouw.edu.pl}

\begin{abstract}

We describe and employ a stacking procedure to investigate 
abundances derived from the low S/N spectra obtained in the Carnegie 
RR Lyrae Survey (CARRS; \citealt{kollmeier13}). 
We find iron metallicities that extend from [Fe/H]~$\sim$~~$-$2.5
to values at least as large as [Fe/H]~$\sim$ $-$0.5 in the 274-spectrum
CARRS RRc data set. 
We consider RRc sample contamination by high amplitude solar metallicity
$\delta$~Scuti stars (HADS) at periods less than 0.3 days, where 
photometric discrimination between RRc and $\delta$~Scuti stars has 
proven to be problematic.  
We offer a spectroscopic discriminant, the well-marked overabundance of 
heavy elements, principally [Ba/H], that is a common, if not universal, 
characteristic of HADS of all periods and axial rotations. 
No bona fide RRc stars known to us have verified heavy-element 
overabundances.
Three out of 34 stars in our sample with [Fe/H]~$>$ $-$0.7 exhibit 
anomalously strong features of Sr, Y, Zr, Ba, and many rare earths.
However, carbon is not enhanced in these three stars, and we conclude 
that their elevated $n$-capture abundances have not been generated in
interior neutron-capture nucleosynthesis.
Contamination by HADS appears to be unimportant, and metal-rich 
RRc stars occur in approximately the same proportion in the Galactic field 
as do metal-rich RRab stars. 
An apparent dearth of metal-rich RRc is probably a statistical fluke.
Finally we show that RRc stars have a similar inverse period-metallicity 
relationship as has been found for RRab stars.

\end{abstract}


%%%%%%%%%%%%%%%%%%%%%%%%%%%%%%%%%%%%%%%%%%%%%%%%%%%%%%%%%%%%%%%%%%%%%%%%%%
\section{INTRODUCTION}\label{intro}
%%%%%%%%%%%%%%%%%%%%%%%%%%%%%%%%%%%%%%%%%%%%%%%%%%%%%%%%%%%%%%%%%%%%%%%%%%

The apparent paucity of metal-rich ([Fe/H]~$>$~$-$1.0) RRc stars in the 
Galactic field has received passing mention during the past few decades, 
viz., \cite{smith95}. 
Only DH Peg, one of the several possible metal-rich RRc candidates in an 
early spectroscopic survey \citep{preston59} survived subsequent photometric 
reclassification, and its membership is debated \citep{fernley90}. 
We are aware of no other bona fide metal-rich RRc star in extant literature. 
None of the twenty RRc in the Hipparcos catalog, chosen primarily by 
proximity to the Sun, are metal-rich \citep{feast08}, nor are any of the 
nineteen RRc chosen for abundance analysis solely by location in the sky at 
Las Campanas Observatory \citep{sneden17}. 
However, such stars ought to exist: theoreticians happily compute 
families of low mass, metal-rich, core helium-burning first 
overtone pulsators, \ie, metal-rich RRc stars \citep{marconi15}.
This is a puzzle that as yet does not have a satisfactory resolution.

The data set of \cite{kollmeier13} (hereafter in this paper K13) provides a 
provisional answer in the form of several dozen metal-rich RRc candidates.  
These stars form the observational basis of the present investigation in which 
we confront two issues. 
\textit{(1)} Low signal-to-noise ($S/N$) of individual stellar echelle spectra, 
the price paid for the rich harvest of the metal-rich candidates, renders 
conventional abundance analysis problematical. 
\textit{(2)} Possible contamination of the K13 metal-rich RRc 
sample by high amplitude $\delta$~Scuti stars (HADS) and/or other related 
members of the near-main-sequence zoo of pulsating stars 
(\citealt{rodriguez00}; see discussion of their Table~12), all of 
which possess more or less solar abundances.
DH Peg illustrates the problem.  
Antonello et al. (1986, \S5)\nocite{antonello86} report, with particular 
reference to DH Peg, that Fourier decomposition of light curves provides 
no clear distinction between HADS and RRc stars.  
Following \citeauthor{antonello86}, Fernley et al. (1990, 
\S4.1)\nocite{fernley90} conclude their analysis of DH Peg with the remark 
that ``DH Peg is most likely an RR Lyrae, but the classification has to be 
treated with some caution''.

We address possible contamination of the K13 sample by $\delta$~Sct stars 
that abound in the solar neighborhood: 636 are listed in the catalogue 
of \cite{rodriguez00}.  
Of these, 54 have $P$~$>$~0.20~d and of these 26 (4\%) are HADS, having light 
amplitudes $\Delta V$~$>$~0.3~mag, values permitted for inclusion in the 
K13 survey. 
Therefore we need (and have found) an effective RRc/$\delta$Sct 
discriminant as we will describe below.

We reconsider the K13 274-spectrum RRc sample in \S\ref{spectra} and 
\S\ref{stackanal}, 
demonstrating that their low $S/N$ spectra yield reliable 
overall [Fe/H] metallicities.
In \S\ref{delscuti} we show that three program stars have significant 
overabundances of neutron-capture ($n$-capture, atomic number $Z$~$>$~30)
elements, and identify these as probable $\delta$~Sct variables.
Finally, in \S\ref{discuss} we 
validate the existence of an anticorrelation between 
[Fe/H] and pulsational period $P$ in RRc stars
analogous to the well-known anti-correlation for RRab stars.

%%%%%%%%%%%%%%%%%%%%%%%%%%%%%%%%%%%%%%%%%%%%%%%%%%%%%%%%%%%%%%%%%%%%%%%%%%
\section{THE SPECTROSCOPIC DATASET}\label{spectra}
%%%%%%%%%%%%%%%%%%%%%%%%%%%%%%%%%%%%%%%%%%%%%%%%%%%%%%%%%%%%%%%%%%%%%%%%%%

The K13 sample began with the RRc variables identified in the 
All Sky Automated Survey catalog (ASAS; \citealt{pojmanski02}, 
\citealt{pojmanski05} and references therein).\footnote{ 
http://www.astrouw.edu.pl/asas/}
\cite{szczygiel07} and \cite{szczygiel09} examined the ASAS RRc light
curves for evidence of the Blazhko effect (a slow variation in photometric and 
spectroscopic periods and amplitudes of some RR Lyrae stars).
The K13 star list did not include any Blazhko variables found in those papers.
Proper motions were taken from USNO CCD Astrograph Catalogs 
\citep{zacharias04,zacharias13}.
High-resolution spectroscopy of candidate RRab and RRc stars was undertaken
to provide accurate radial velocities and approximate values of metallicities.
K13 considered only RRc stars, with the RRab sample to be considered later.

The spectra for the K13 snapshot RR~Lyrae survey were gathered in 2011-12 
with the echelle spectrograph of the Las Campanas Observatory 2.5m du Pont 
Telescope.
The spectrograph configuration was identical to that employed in a series
of papers on the velocities, H$\alpha$ profiles, metallicities, and 
relative abundance ratios of RR~Lyrae stars (\citealt{for11a,for11b}, 
\citealt{govea14}, \citealt{chadid17}, \citealt{sneden17}).
The spectrograph with a 1.5$\times$4.0\,arcsec aperture delivered resolving power
$R$~=~$\lambda/\Delta\lambda$~$\sim$~27000 at 5000\,\AA.
The spectral coverage was $\lambda\lambda$ 3400--9000, with
wavelength gaps in the CCD order coverage beginning at $\lambda$~7100~\AA\
and growing with increasing wavelength.
In practice the useful spectral domain for this work was limited to 
$\lambda\lambda$ 3900--7000~\AA.

The K13 spectra were taken as close as possible to a target pulsational phase
$\phi$~=~0.32, which is the phase of the time-averaged 
velocity on the ascending branch of the RV curve (with respect to visual 
maximum light defined as $\phi$~=~0) for the RRc variables T~Sex and TV~Boo
\citep{liu89}, DH~Peg \citep{jones88a}, and YZ~Cap \citep{govea14}.\footnote{
The time-average RV of RRab stars is $\phi$~$\simeq$~0.37 
(\eg, \citealt{liu91}, \citealt{preston11}).}.
The phase and radial velocity issues are discussed in \S3 of K13.
The target phase was achieved in their 261-star RRc sample, as illustrated in
K13 Figure~3; we calculate $\langle\phi\rangle$~=~0.333 ($\sigma$~=~0.024).
The reduction path from observed CCD frames to final wavelength-calibrated
multi-order spectra used a software pipeline designed 
and implemented by \cite{kelson03}.
Continuum-normalized and order-concatenated spectra were produceed with the
IRAF ECHELLE\footnote{
\cite{tody93}; IRAF is distributed by the National Optical Astronomy 
Observatory, which is operated by the Association of Universities for 
Research in Astronomy, Inc., under a cooperative agreement with the National 
Science Foundation.} 
package.
The observed RRc sample is given in Table~\ref{tab-stars}, identified by their
ASAS number.
This table is ordered according to derived K13 metallicity, with the
highest first (124115--4056.9, [Fe/H]~=~$+$0.04) and the lowest last
(195307-5131.1, [Fe/H]~=~$-$2.78)
Among the listed quantities taken from the ASAS database are 
$V_{max}$, the mean visual magnitude at maximum light and $V_{amp}$, the 
visual light amplitude (the defined zero point of pulsational phase $\phi$). 
There were 274 observed spectra, and with 13 stars observed twice, the
total number of stars was 261.
In the present paper we treat individual spectra as independent objects.

The K13 metallicity estimates were made by matching the observed spectra to
a grid of synthetic spectra, as described in their \S3.3.
RRc stars have well-determined photometric properties with relatively 
small variations throughout their pulsational cycles.  
Mean absolute magnitudes are $\langle M_V\rangle$~=~$+$0.6~$\pm$~0.1 (K13)
and cyclical variations are $\langle M_{amp}\rangle$~$\simeq$~0.25 
(Table~\ref{tab-stars}).
Similarly, $\langle (B-V)_{amp}\rangle$~$\sim$~0.2 (\eg, Figure~3 of
\citealt{layden13}).
The small variations in these photometric quantities combined with the
very small observed pulsational phase range centered at $\phi$~=~0.33
allowed K13 to adopt a single set of atmospheric parameters except
overall metallicity, based on their analysis of higher $S/N$ du Pont echelle
spectra of the RRc star YZ~Cap (not included in the K13 sample).
The values derived for YZ~Cap were effective temperature \teff~=~7000~K, 
surface gravity \logg~=~2.2, microturbulent velocity \vmicro~= 2.5~\kmsec, and
$\alpha$ element enhancement [$\alpha$/Fe]~=~+0.35.  

The wavelength regions studied by K13 were limited to a 
``blue'' spectral interval 4400$-$4675~\AA\ and a ``yellow'' interval 
5150$-$5450~\AA.  
The chosen yellow interval contains many neutral-species metal transitions, and
especially includes the \species{Mg}{i} b lines; this triplet remains easily
detectable even in the most metal-poor RRc target.  
The majority of strong lines in the blue interval are due to 
metal ions, such as \species{Ti}{ii} and \species{Fe}{ii}.
This region also contains the \species{Ba}{ii} 4554~\AA\ line that is often
used to identify potential neutron-capture-rich stars, and proved to be of
importance to the present investigation.
Other spectral regions were ignored, for various reasons.
Briefly, the observed spectra extend significantly blueward of 4400~\AA\ and 
they display many useful spectral absorption features.
Unfortunately, the $S/N$ of these snapshot spectra declines 
substantially below 4400~\AA, rendering this spectral region less useful for 
metallicity estimates.
The yellow-red regions beyond 5450~\AA\ have adequate $S/N$, but our
warm, metal-poor RRc stars have very few strong lines in this wavelength
domain; many intervals appear to be close to pure continua.
The region between 4675 and 5150~\AA\ has a few potentially useful lines,
but it includes H$\beta$ at 4861~\AA, whose broad wings 
complicate abundance analysis.

In K13 the line lists for synthetic spectra began with the \cite{kurucz11} 
database\footnote{
http://kurucz.harvard.edu/}
followed by syntheses of the solar spectrum in order to adjust the line
parameters (mostly their transition probabilities).
Model atmospheres were interpolated among the \citeauthor{kurucz11} ATLAS
model grid, and synthetic spectra were computed with the current version
of the MOOG synthetic spectrum code \citep{sneden73}\footnote{
Available at http://www.as.utexas.edu/~chris/moog.html}.
The model grid was computed from $+$0.05~$\geq$~[Fe/H]~$\geq$~$-$2.90.
K13 then used reduced $\chi^2$ minimization (weighting by $S/N$) to 
estimate [Fe/H] values separately in the blue and yellow spectral regions. 
Within each region estimates were made covering the whole interval and one 
more narrowly confined to a set of the strongest lines.  
Thus four metallicity estimates were made for each RRc star.

In Table~\ref{tab-stars} for each star we list the mean $S/N$ of the values in 
the blue and yellow spectral intervals, and the mean K13 [Fe/H] 
derived from the four individual estimates.
A simple average yields $\langle S/N\rangle$~=~18.5 
($\sigma$~=~6.6)\footnote{
The values in blue and yellow spectral regions (not given in 
Table~\ref{tab-stars}) are $\langle S/N\rangle$~=~15.1 and 
$\langle S/N\rangle$~=~22.0, respectively.}.
The average scatter of the four [Fe/H] estimates leading to the final 
values in Table~\ref{tab-stars} is $\langle\sigma\rangle$~=~0.18; this is
a reasonable estimate of internal uncertainties in the K13 metallicity
values.

%%%%%%%%%%%%%%%%%%%%%%%%%%%%%%%%%%%%%%%%%%%%%%%%%%%%%%%%%%%%%%%%%%%%%%%%%%
\section{NEW RRC METALLICITIES FROM CO-ADDED SPECTRA}\label{stackanal}
%%%%%%%%%%%%%%%%%%%%%%%%%%%%%%%%%%%%%%%%%%%%%%%%%%%%%%%%%%%%%%%%%%%%%%%%%%

The low $S/N$ of the K13 spectra precludes the kind of line-by-line model 
parameter and abundance analysis done in our previous RR~Lyrae studies cited
above.
Given the narrow range of RRc stellar atmosphere quantities and the large 
number of stars in all metallicity domains, we decided to co-add the K13
spectra in small metallicity bins in order to significantly increase their
$S/N$ so that a more traditional investigation could be undertaken.

%%%%%%%%%%%%%%%%%%%%%%%%%%%%%%%%%%%%%%%%%%%%%%%%%%%%%%%%%%%%%%%%%%%%%%%%%%
\subsection{Stacking the Spectra}\label{makestack}
%%%%%%%%%%%%%%%%%%%%%%%%%%%%%%%%%%%%%%%%%%%%%%%%%%%%%%%%%%%%%%%%%%%%%%%%%%

Beginning at the high-metallicity end of the K13 program stars, we averaged 
successive sets of 10 spectra, weighting the means by the $S/N$ of 
each spectrum.
We will refer to these mean spectra as ``stacks'' and their 
properties (metallicity, temperature, etc.) as ``stack" properties.
With 274 program stars, there were 28 stacks, with the lowest
metallicity stack\#28 continuing just four spectra.  
Additionally, for reasons to be developed below, we believe that three
stars in stack\#1 are not members of the RRc class.
These stars are labeled ``no'' in the last column of Table~\ref{tab-stars}.
Since these three do not appear to be true RRc stars, 
this stack contains only seven spectra.
We list the 28 stacks in Table~\ref{tab-means}. 
For each stack we give the number of stars, the mean K13 [Fe/H] values, the 
spread of the K13 metallicity values $\Delta$[Fe/H], model atmosphere 
parameters (see \S\ref{stackmods}), and the mean pulsational periods.
As illustrated in the K13 Figure~1 metallicity histogram, there
are comparatively few stars at the high and low metallicity ends of our
sample, and a large pile-up near [Fe/H]~$\simeq$~$-$1.2; 
this is reflected in
the different $\Delta$[Fe/H] values in Table~\ref{tab-stars}.

\begin{figure}
\epsscale{1.00}
\plotone{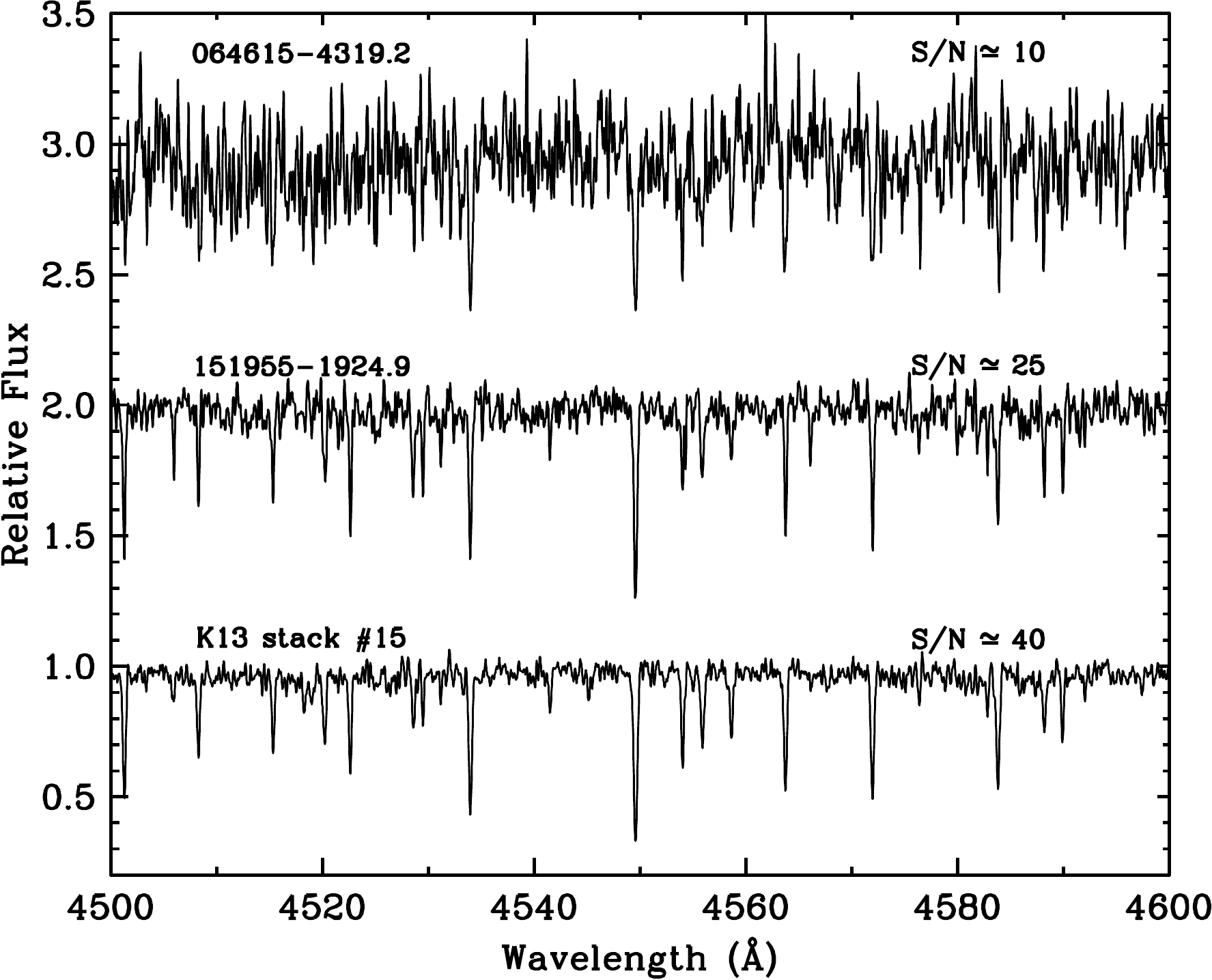}
\caption{\label{f1}
\footnotesize
   A small part of the short-wavelength interval in the K13 10-star stack 
   \#15 (lower spectrum), and in two of the stars in this stack.
   Star 151955-1924.9 (middle spectrum) has the highest $S/N$ in the
   stack, and 064615-4319.2 (top spectrum) as the lowest $S/N$ in the stack.
}
\end{figure}

In Figure~\ref{f1} we show a typical example of the improvement in 
$S/N$ achieved by the stacking procedure.  
A small part of the K13 blue spectral region has been chosen to illustrate
a difficult $S/N$ regime.
The stack\#15 co-added spectrum ([Fe/H]$_{K13}$~=~$-$1.26) is shown
along with two of its constituent spectra that were chosen to represent the
highest and one of the lowest $S/N$ values.
For stack\#15 in the blue we estimate $S/N$~$\sim$~40 as indicated in 
the figure, and in the yellow $\sim$~60; values for other stacks 
are comparable.

%%%%%%%%%%%%%%%%%%%%%%%%%%%%%%%%%%%%%%%%%%%%%%%%%%%%%%%%%%%%%%%%%%%%%%%%%%
\subsection{Stack Analyses}\label{stackmods}
%%%%%%%%%%%%%%%%%%%%%%%%%%%%%%%%%%%%%%%%%%%%%%%%%%%%%%%%%%%%%%%%%%%%%%%%%%

We treated each stack as an observed stellar spectrum, and derived
atmospheric quantities in the same manner as done by \cite{chadid17}, 
\cite{sneden17}, and references therein.
The line analysis code and model atmosphere databases were as described
in \S\ref{spectra}.
However, unlike the K13 linelists developed by matching the solar
spectrum in two regions, for this work we expanded the spectral range to 
3900$-$6500~\AA\ and used only unblended absorption lines with laboratory 
transition probabilities.  
Because the present study concentrates on Fe metallicities and just a few 
indicators of elemental group relative abundances, we list only the most 
important $gf$-value sources (see \citealt{sneden16} for extended comments).  
$(a)$ \species{Fe}{i}: \cite{obrian91} for lines with excitation energies
$\chi$~$<$~2.4~eV and \cite{denhartog14}, \cite{ruffoni14} for higher
excitation lines.
$(b)$ \species{Fe}{ii}: the NIST Atomic Spectra Database\footnote{
http://physics.nist.gov/PhysRefData/ASD/} \citep{NIST15}.
$(c)$ \species{Ti}{ii}: \cite{wood13}; a few \species{Ti}{i} lines were also
measured, and the $gf$'s of \cite{lawler13} were used, but they provided
only a weak check on the Ti abundances derived from ionized lines.

Standard excitation equilibrium constraints and weak-line/strong-line
balance on \species{Fe}{i} transitions were used to iteratively derive 
\teff\ and \vmicro, respectively.
Ionization balance between \species{Fe}{i} and \species{Fe}{ii} yielded
\logg, and derived [Fe/H] values were checked for consistency with input
model metallicities.
We list these quantities for the stacks in columns 5$-$8
of Table~\ref{tab-means}.
The variations in these quantities over the complete set of stacks
(\eg, over the whole RRc metallicity range) are modest:
$\langle\teff\rangle$~=~7015, $\sigma$~=~90~K;
$\langle\logg\rangle$~=~2.6, $\sigma$~=~0.3; and
$\langle\vmicro\rangle$~=~2.7, $\sigma$~=~0.5.

More importantly, the newly-derived stack [Fe/H] values correlate well 
with the mean K13 stack metallicities.
Taking differences in the sense 
$\Delta$[Fe/H]~$\equiv$ [Fe/H]$_{K13}$~$-$~[Fe/H]$_{new}$,
the average is $\langle\Delta$[Fe/H]$\rangle$~=~$-$0.11~$\pm$~0.03
($\sigma$~=~0.17).
However, there is a small, nearly linear drift with metallicity in 
the $\Delta$[Fe/H] values, as we illustrate in Figure~\ref{f2}.
A linear regression line is [Fe/H]$_{new,pred}$ = 0.75[Fe/H]$_{K13}$~$-$~0.44
with correlation coefficient $r$~=~0.98.
The scatter of points around this trend line is very small, $\sigma$~=~0.08.

Overall we regard comparison between our new stack metallicities and those
of K13 as very good.
There are a number of possible causes for the slope of the regression seen 
in Figure~\ref{f2}; here we list a few of the more likely explanations.
First, continuum normalizations are difficult to do self-consistently at
the low $S/N$ of the original K13 spectra.  
This is caused by the more than two dex metallicity range of the stellar
sample, which created large differences in line density in both the original
blue and yellow spectral regions.
Second, the K13 $\chi^2$ metallicity estimation technique was fairly simple. 
In that study it was easiest to find reliable $\chi^2$ minima in the 
higher metallicity regime.
Third, the new analysis relies on laboratory transition data for individual
spectral lines; K13 based their spectrum syntheses on line lists that
matched the solar spectrum.
However, the basic result here is that the K13 metallicities correlate
well with standard abundance analyses; the grid synthesis method yielded 
reliable metallicities even when the spectral $S/N$ values were very low.

In Figure~\ref{f3} we compare two 
stacked spectra with those of two individual RRc stars from 
\nocite{sneden17}Sneden et al. (2017, hereafter S17), displaying a portion of 
the wavelength region that has the prominent \species{Mg}{i} b line triplet. 
That study observed 19 RRc stars throughout their pulsational cycles, 
combining new data with those published by \cite{govea14}.
Typically about 40 spectra per star were gathered, enough so that multiple 
observations at one phase interval ($\delta\phi$~$\simeq$~$\pm$0.05) 
could be combined to achieve fairly high $S/N$. 
Here we consider two stars at the extremes of the \citeauthor{sneden17} 
metallicity range. 
The S17 RRc stars were analyzed both from spectra at individual phases 
and from co-additions into stellar spectrum means.
Star 190212-4639.2 (also part of the K13 survey) was analyzed in
various ways, leading to [Fe/H]~=~$-$2.65, with probable uncertainty 
$\simeq\pm$0.15.
That star is a member of stack\#28, which has [Fe/H]$_{K13}$~=~$-$2.69 
and [Fe/H]$_{new}$~=~$-$2.42 (Table~\ref{tab-means}).
For Figure~\ref{f3} we have displayed the 
pulsational phase $\phi$~=~0.331 S17 spectrrum, for which they derive
[Fe/H]~=~$-$2.62 with $\sigma$~= 0.16.
This spectrum is virtually indistinguishable from that of stack\#28 
in the figure.
At the high metallicity end, we display the 
S17 spectrum of 132226-2042.3 (with [Fe/H]~=~$-$0.95, probable uncertainty 
$\simeq\pm$0.15) and that of our stack\#5 ([Fe/H]$_{K13}$~=~$-$0.80,
[Fe/H]~=~$-$1.03).
We have chosen to show their $\phi$~=~0.448 spectrum, for which they 
derive [Fe/H]~=~$-$1.03.
Again, visual comparison of the spectra in Figure~\ref{f3}
suggests excellent agreement.

We have not made extensive chemical composition analyses of the stacked 
spectra for which the $S/N$ values remain modest after stacking.
However, usually there are enough \species{Ca}{i} and \species{Ti}{ii} 
lines to check relative abundances in the $\alpha$ element group.
For the 28 stacks we derive $\langle$[Ca/Fe]$\rangle$~=~$+$0.25 
($\sigma$~=~0.13) and $\langle$[Ti/Fe]$\rangle$~=~$+$0.50 ($\sigma$~=~0.20).
These reproduce, with large scatter especially for Ti, the well-known
$\alpha$-element overabundances in metal-poor stars.  
We also note that for stack\#1, our highest metallicity spectrum,
we derive [Ca/Fe]~$\simeq$~ [Ti/Fe]~$\simeq$~0.0, as expected for relatively
metal-rich stars.
More detailed abundance analyses are not justified here.

%%%%%%%%%%%%%%%%%%%%%%%%%%%%%%%%%%%%%%%%%%%%%%%%%%%%%%%%%%%%%%%%%%%%%%%%%%
\section{$\delta$ SCUTI STARS HIDING IN PLAIN SIGHT}\label{delscuti}
%%%%%%%%%%%%%%%%%%%%%%%%%%%%%%%%%%%%%%%%%%%%%%%%%%%%%%%%%%%%%%%%%%%%%%%%%%

As outlined in \S\ref{intro} metal-rich RRc stars are 
difficult to distinguish from high amplitude $\delta$~Scuti stars (HADS) 
by multi-color photometry. 
Fortunately a clear spectroscopic signal exists: (relatively) slowly
rotating $\delta$~Scuti stars have enhanced abundances of heavy elements, 
generally rising with increasing atomic number to become extremely high in the 
$n$-capture domain ($Z$~$>$~30).
We illustrate this by gathering literature [Fe/H] and [Ba/H] 
values for HADS and plotting them in Figure~\ref{f4}.
Panel (a) of this figure shows that the Fe abundances have some star-to-star
scatter, but with a nearly solar mean:  $\langle$[Fe/H]$\rangle$~=~$+$0.12 
($\sigma$~=0.20, 23 stars).
Several stars with $v sin i$~$<$~100~\kmsec\ have [Fe/H]~$>$~$+$0.2,
which may be a sign of the general elevation of heavy element overabundances
of $\delta$~Scuti stars.  
However, relative abundances of Ba ($Z$~= 56) displayed in 
panel (b) are all supersolar, all having [Ba/Fe]~$\gtrsim$~$+$0.2.
The majority of these $\delta$~Scuti's have [Ba/H]~$\gtrsim$~0.5, and six have 
have [Ba/H]~$\gtrsim$~1.  

We searched the higher metallicity K13 spectra for signs of $n$-capture
overabundances.
In spite of the low $S/N$ of these spectra, three stars were easily spotted
with unusually strong \species{Ba}{ii} 5853~\AA\ lines:  ASAS
095845-5927.8, 195123+0835.0, and 141107-4212.2 (tagged with ``no'' 
entries in Table~\ref{tab-stars}).
This is significant because normally this line cannot be easily detected 
in the K13 spectra even in the metal-rich domain.
Therefore, we formed a mean spectrum of these three Ba-strong stars and 
subjected it to the same analysis as described above for the stack spectra.
We derived model parameters \teff~=~6900~K, \logg~=~3.0, 
\vmicro~=~3.2~\kmsec, and [Fe/H]~=~$-$0.45.
These values are similar to those of stack\#1 except for the gravity:
\logg\ is greater by 0.4~dex in the Ba-strong mean spectrum.

The heavy element overabundances of Ba-strong stars are obvious by 
inspection of relevant transitions. 
Figure~\ref{f5} presents the spectroscopic evidence. 
Panels (a) and (b) show the \species{Ba}{ii} 5853~\AA\ and 6141~\AA\ lines of 
stack\#1 and those of a co-addition of the three Ba-strong stars excluded 
from this stack.  
Significant Ba strength differences are evident, most
clearly for the 5853~\AA\ line.
The 6141~\AA\ feature is partially blended with an \species{Fe}{i} line,
lessening the total difference.
These sharp contrasts in \species{Ba}{ii} strengths 
are shared by very strong 4554~\AA, strong-but-blended 4934~\AA, 
clean and strong 6496~\AA, and high excitation weak-but-blended 4130~\AA.
We computed synthetic spectra of the 4554, 5853, 6141, and 6496~\AA\ lines,
deriving $\langle$[Ba/H]$\rangle$~=~$-$0.58 ($\sigma$~=~0.16) for stack\#1 and
$\langle$[Ba/H]$\rangle$~=~$+$1.00 ($\sigma$~=~0.15) for the Ba-strong spectrum.
The 1.6~dex overabundance in the Ba-strong stars confirms the \species{Ba}{ii}
line strength differences seen in Figure~\ref{f5}.
However, much caution is warranted on the magnitude of the overabundance, 
because nearly all \species{Ba}{ii} lines are very 
saturated in the Ba-strong stars. 
Therefore they are sensitive to adopted \vmicro\ and to outer-atmosphere 
line formation effects that are not accounted for in our standard 
modeling approach.
Such very strong lines typically yield unrealistically large abundances
of Fe-group elements in RRc stars, and thus are ignored in our model analyses.

Fortunately other heavy $n$-capture elements are easily detected in our
Ba-strong stars: La ($Z$~=~57), Eu (63), and to a lesser extent Nd (60).
Six lines of \species{La}{ii} are seen, and the most useful ones at 
4123, 4920, and 4921~\AA\ yield for the mean three-star Ba-strong spectrum 
$\langle$[La/H]$\rangle$~$\simeq$~$+$0.2, while for the stack\#1 spectrum 
we obtain $\langle$[La/H]$\rangle$~$\simeq$~$-$0.4.
Lines of \species{Eu}{ii} at 4129 and 4205~\AA\ suggest
$\langle$[Eu/H]$\rangle$~$\simeq$~$+$0.4 for the Ba-strong stars while
they are undetectable in the stack\#1 spectrum.
Taken together, the three heavy $n$-capture elements have 
[X/H]~$\sim$~$+$0.2 to 1.0,
or [X/Fe]~$+$0.7 to $+$1.5 in the Ba-strong spectrum while the stack\#1 
spectrum has [X/H]~=~$-$0.6 to $-$0.4 or [X/Fe]~=~~$-$0.1 to $+$0.1.

The overabundances extend to the lighter $n$-capture elements Sr, Y, and Zr 
($Z$~=~38$-$40).
Panels (c) and (d) of Figure~\ref{f5} show that the Ba-strong stars 
exhibit much stronger \species{Y}{ii} lines than those of stack\#1.
From these lines we derive $\langle$[Y/H]$\rangle$~$\simeq$~$+$0.2, or
$\langle$[Y/Fe]$\rangle$~$\simeq$~$+$0.6 in the Ba-strong stars, while
two lines of \species{Zr}{ii} in the noisy and crowded 4200~\AA\ spectral
domain yield $\langle$[Zr/H]$\rangle$~$\simeq$~$+$0.4, or
$\langle$[Y/H]$\rangle$~$\simeq$~$+$0.8.
We did not attempt syntheses of the extremely strong \species{Sr} 4077 
and 4215~\AA\ resonance lines, but inspection of our spectra argues for
the same abundance enhancement in Ba-strong stars for Sr as well.

In contrast, the evolutionary-sensitive light element C shows no abundance 
difference between the Ba-strong spectrum and that of stack\#1.
In fact, the \species{C}{i} high excitation lines ($\chi$~$>$~7.5~eV)
displayed in Figure~\ref{f5} panels (e) and (f) appear to be
weaker in the Ba-strong stars.
This is confirmed by our spectrum syntheses of these two lines plus
\species{C}{i} 5052~\AA: $\langle$[C/H]$\rangle$~$\simeq$~$-$0.3
in the Ba-strong spectrum and 
$\langle$[C/H]$\rangle$~$\simeq$~$-$0.1 in stack\#1.  
The lack of C enhancement in the Ba-strong spectrum is confirmed by our
failed attempt to detect the CH G-band.  
Even at \teff~$\sim$~7000~K, a substantial C overabundance would have
produced measurable CH absorption near the bandheads in the 4300$-$4325~\AA\
region.

In summary, all $n$-capture elements are very overabundant but C is not
in the Ba-strong stars.
Setting aside analytically-difficult Ba, we suggest that
[$\langle$Y,Zr,La,Eu$\rangle$/Fe]~$\simeq$~$\sim$~$+$0.7.  
This cannot be the result of rapid $n$-capture nucleosynthesis (the 
$r$-process), because the signature element of the $r$-process is greatly
enhanced Eu/La values; this is contrary to our result.
However, slow $n$-capture (the $s$-process) cannot be the cause either:
not only is there no La/Eu overabundance, there is no evidence for
enhanced C, which is a general characteristic of $s$-process-rich stars.
We conclude that stellar evolutionary processes cannot be blamed for
the large heavy element abundances in our Ba-strong stars.  
Therefore, we think that these stars are HADS.

%%%%%%%%%%%%%%%%%%%%%%%%%%%%%%%%%%%%%%%%%%%%%%%%%%%%%%%%%%%%%%%%%%%%%%%%%%
\section{CONCLUSIONS AND DISCUSSION}\label{discuss}
%%%%%%%%%%%%%%%%%%%%%%%%%%%%%%%%%%%%%%%%%%%%%%%%%%%%%%%%%%%%%%%%%%%%%%%%%%

The two basic results of this paper are that (1) the K13 
high resolution-but-low $S/N$ spectra of RRc candidates yield reliable 
[Fe/H] metallicities on average for RRc stars, and (2) among the K13 RRc 
sample there are a few high metallicity stars with overabundances of 
$n$-capture elements that identify them as $\delta$~Scuti stars.
Here we consider the new metallicities in more detail.
Following \cite{chadid17} and \cite{sneden17}, we chose Layden's 
(1995)\nocite{layden95a} division between metal-poor (MP) and metal-rich (MR) 
RRL stars at [Fe/H]~=~$-$1. 
The MR sample should be dominated by Galactic thin and thick-disk members,
and the MP sample should contain mostly halo members.
The exact choice of the MP/MR metallicity division is not important for
our discussion.

The K13 RRc metallicities, now confirmed by our 
stacked-spectrum analyses, exhibit a small but statistically significant 
anti-correlation with photometric pulsational periods.  
An inverse relationship between [Fe/H] and period for RRab stars has been 
known for some time. 
\cite{preston59} defined a metallicity-sensitive \species{Ca}{ii} K-line 
strength index $\Delta S$ from low-resolution spectra of about 100 
RRab variables.  
His Figure~4 showed that the highest metallicity stars 
([Fe/H]~$\sim$~0.0, $\Delta S$~$\sim$~0) have periods 
$\langle P\rangle$~$\sim$~0.4~days while the lowest metallicity stars
([Fe/H]~$\sim$~$-$2.5, $\Delta S$~$\sim$~10) have 
$\langle P\rangle$~$\sim$~0.7~days.
\citeauthor{preston59} also suggested that a period-metallicity 
anticorrelation exists for RRc stars, but this conclusion was based on 
only nine such stars in his sample.

In Figure~\ref{f6} we use the data of Table~\ref{tab-means}
to show the metallicity-period relationship for our RRc spectrum stacks, 
plotting points for both the original K13 metallicities and those newly 
derived in this paper.
The metallicity-period anticorrelation is evident in this plot.
However, a few cautions should be borne in mind.
First, note the ``mean stack ranges'' in the plot that is placed in the 
lower left corner.
The dimensions of this cross were formed by calulating the means of 
the spreads of periods and metallicities in each of the 28 stacks.
This is representative, but individual stacks can have larger/smaller 
spreads in either quantity.
The ranges suggest that typical stacks have only small star-to-star scatter in 
[Fe/H] values but large scatter in their pulsational periods.
Second, the points at the extreme metallicity ends should not be 
over-interpreted.
For example, there are just four stars comprising the most metal-poor 
stack\#28, and seven in the most metal-rich stack\#1 (discussed in 
\S\ref{delscuti}).
Third, on our revised metallicity scale, the total number of MR stars 
is only 38 (14\% of our sample).

\begin{figure}
\epsscale{1.00}
\plotone{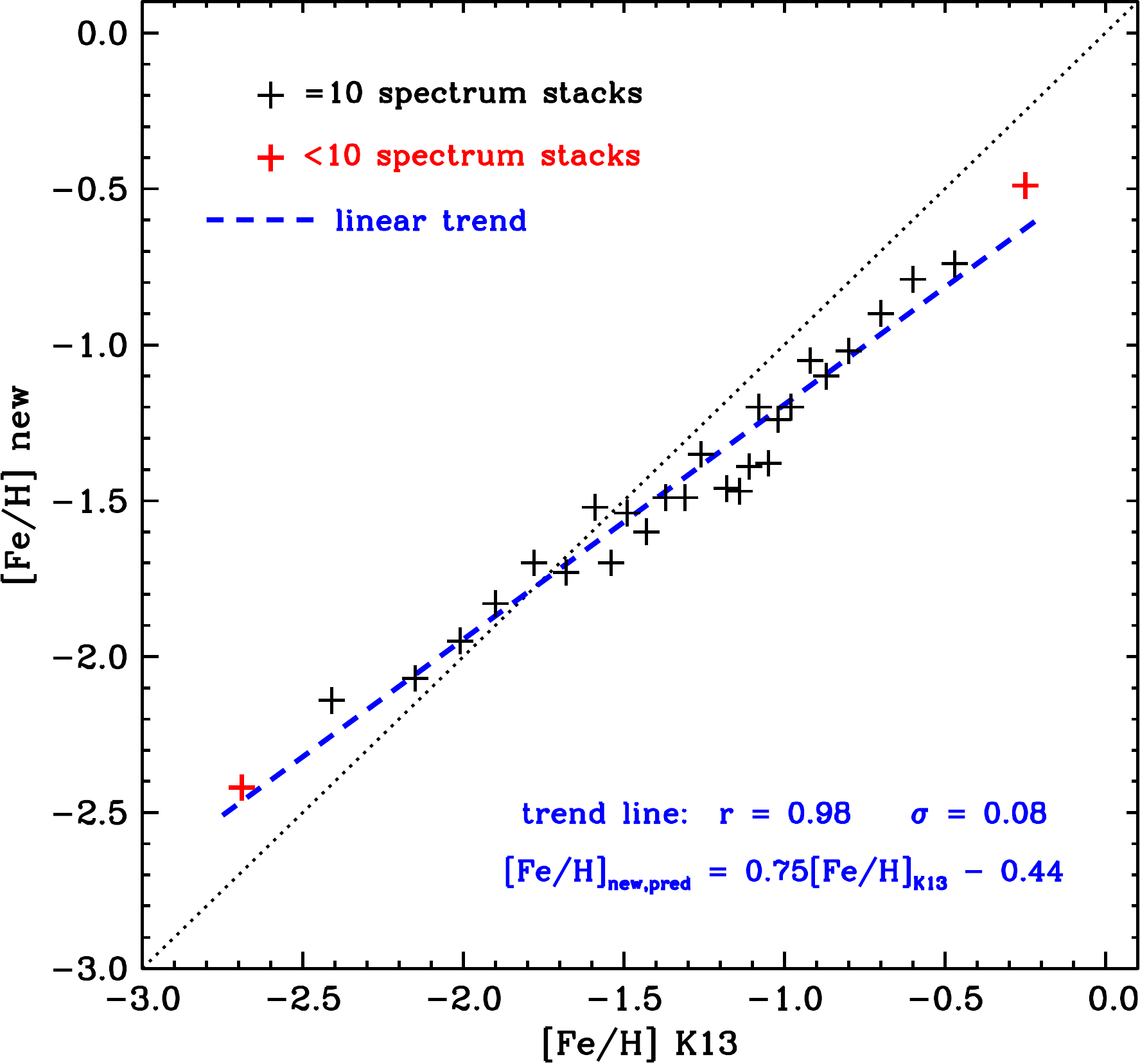}
\caption{\label{f2}
\footnotesize
   Comparison of metallicities from co-added K13 spectra and the mean
   values derived in the original K13 analyses.
   The dotted line represents equality of the [Fe/H] values.
   The blue dashed line represents the linear regression
   line through the data; its parameters are entered in the figure.
}
\end{figure}

\begin{figure}
\epsscale{1.00}
\plotone{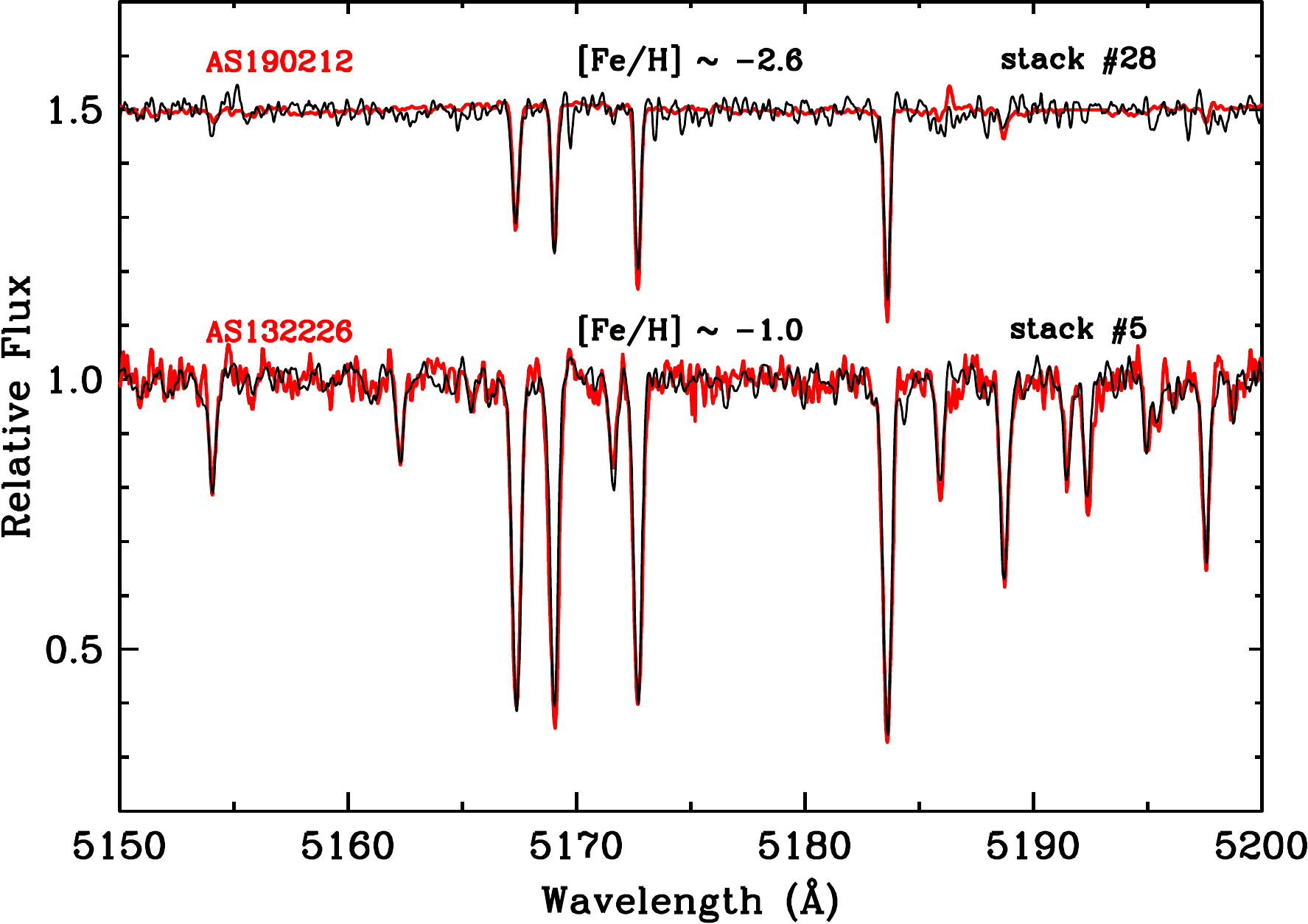}
\caption{\label{f3}
\footnotesize
   Comparison of our stacked spectra (black lines) and ond those
   selected from the \cite{sneden17} RRc study (red lines) in spectral region
   surrounding the \species{Mg}{i} b lines.
   The two examples are among the most metal-poor and metal-rich of the
   RRc study.
}
\end{figure}

\begin{figure}
\epsscale{1.00}
\plotone{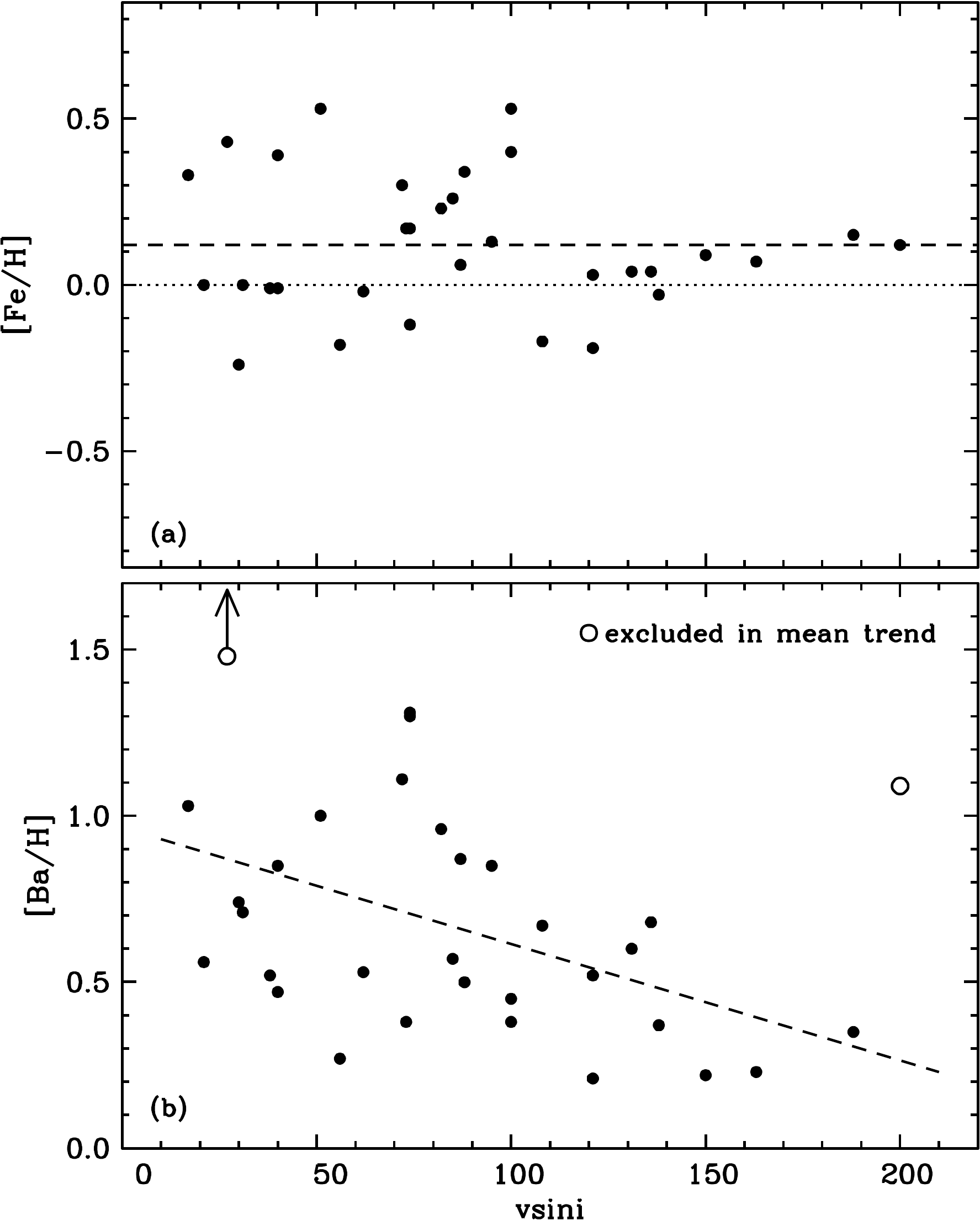}
\caption{\label{f4}
\footnotesize
   Correlations of [Fe/H] metallicity (panel a) and [Ba/H] abundances
   (panel b) in $\delta$ Scuti stars.  
   In panel (a) the dotted line represents the solar value, [Fe/H]~=~0,
   and the dashed line represents the mean of the data points, [Fe/H]~=~$+$0.12.
   In panel (b) the dashed line shows a linear regression through the data
   with the exclusion of the anomalous values 
   for HD~176643 ($v$sin$i$~=~27~\kmsec, [Ba/H]~=~2.2)
   and HD~73798 ($v$sin$i$~=~200~\kmsec, [Ba/H]~=~1.0).
   These data were taken from \cite{ishikawa75}, \cite{mittermayer03}, 
   \cite{yushchenko05}, \cite{fossati08a,fossati08b}, \cite{balona11},
   \cite{catanzaro14}, \cite{escorza16}, and \cite{joshi17}.
}
\end{figure}

\begin{figure}
\epsscale{1.00}
\plotone{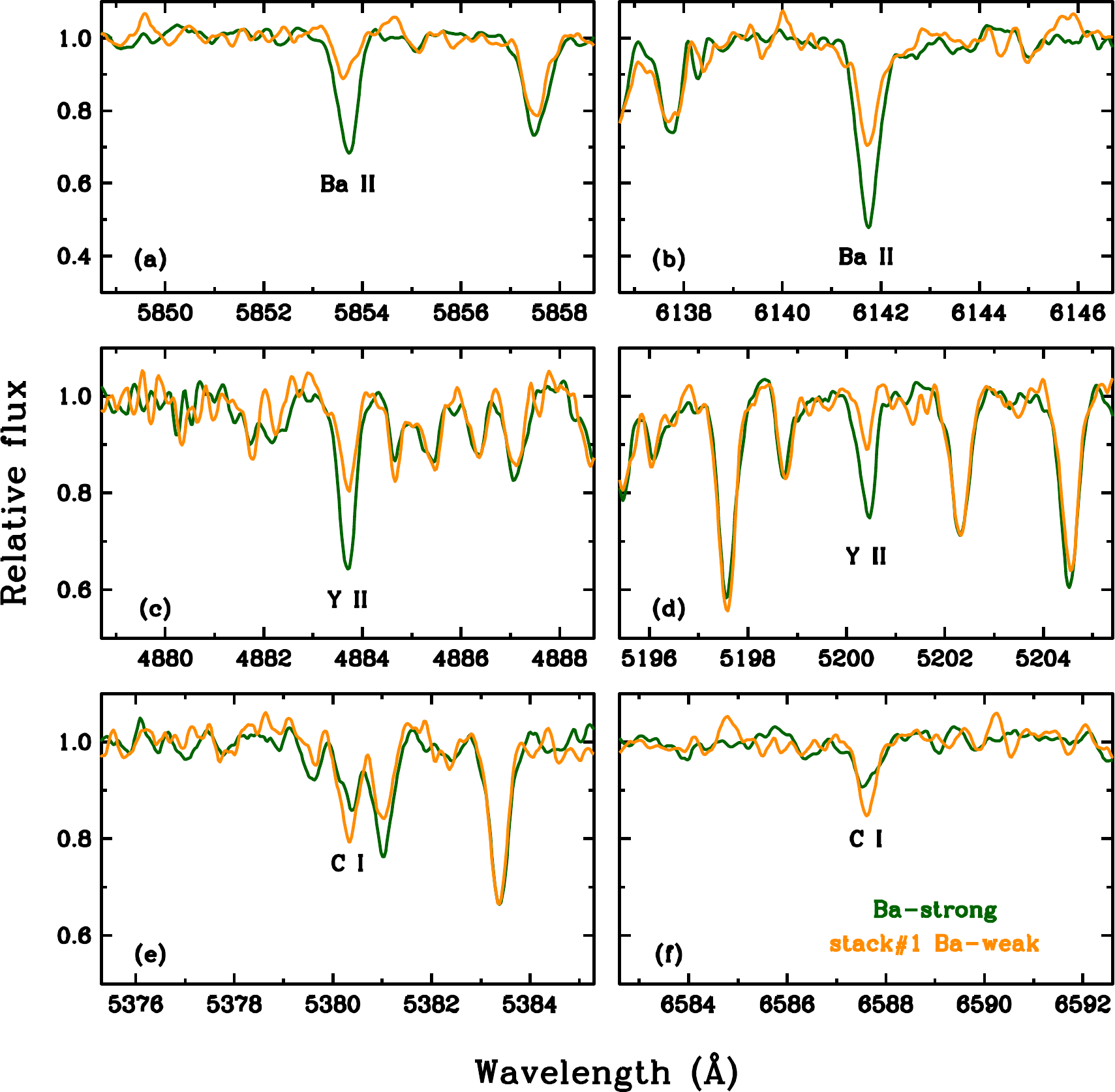}
\caption{\label{f5}
\footnotesize
   Six small spectral windows chosen to display features of Ba, Y, and C
   in stack\#1 (orange lines) and Ba-strong (green lines) spectrum.
   As described in the text, the Ba-strong spectrum is the mean of the
   spectra that were excluded from stack\#1 due to their high $n$-capture
   line strengths.
}
\end{figure}

\begin{figure}                                               
\epsscale{1.00}                                              
\plotone{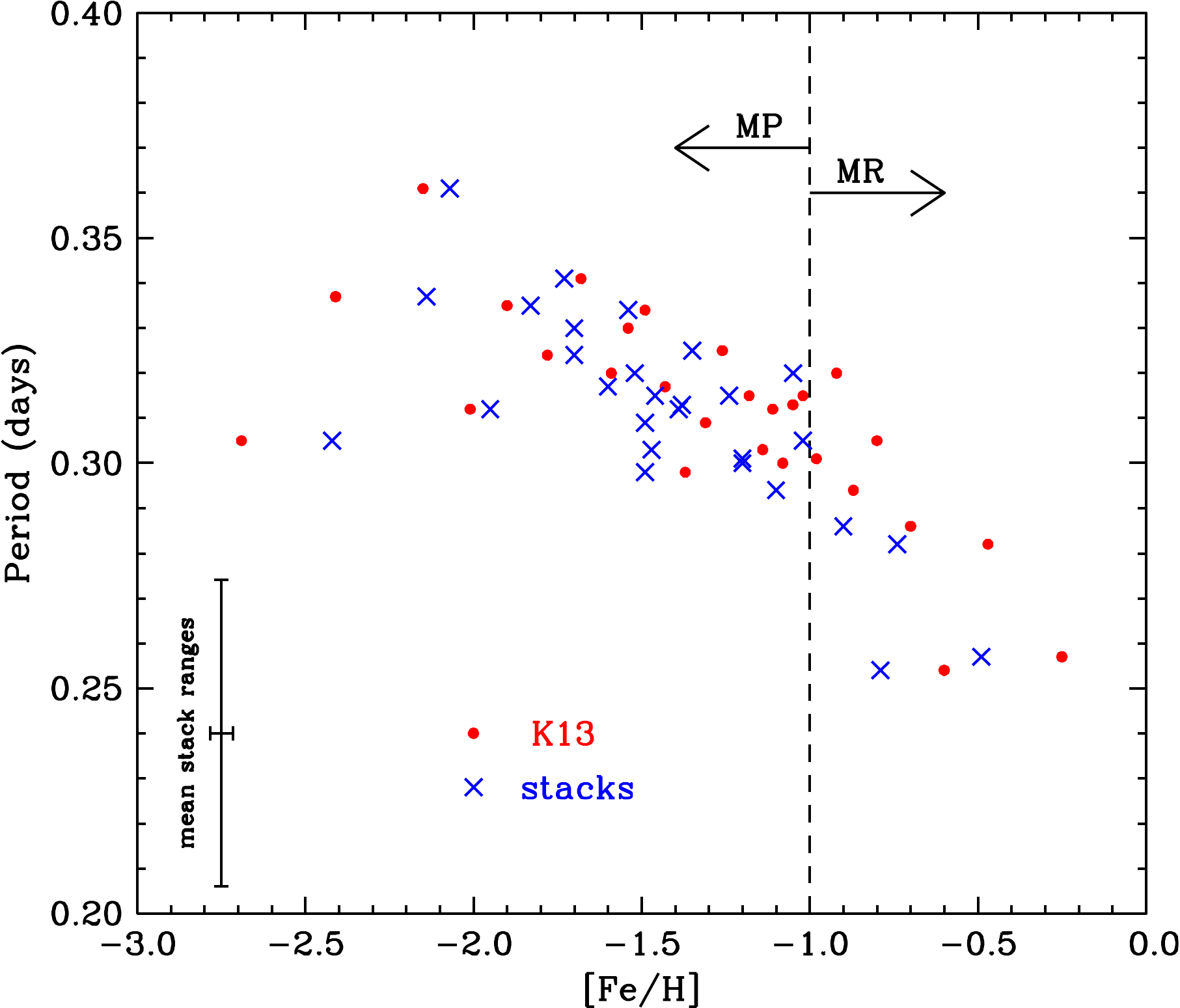}                                         
\caption{\label{f6}                                      
\footnotesize                                                
   Pulsational period plotted as a function of [Fe/H] metallicity
   for our stacked spectra.
   The red dots are K13 values and the blue crosses are from the present
   study (Table~\ref{tab-means}. 
   The black dashed line represents the division adpoted in the paper betwee 
   metal-rich (MR) and metal-poor (MP) stars.
   The mean stack ranges drawn with black solid lines represents mean ranges
   of [Fe/H] and period in a single stack.
}                                                            
\end{figure}

With these caveats in mind, the metallicity-period trend for 
RRc stars is clear in Figure~\ref{f6} whether the original K13 stack 
[Fe/H] values or the ones derived in this paper are used.
The new metallicities and those of K13 agree at [Fe/H]~$\sim$~$-$1.8,
with the new scale yielding higher [Fe/H] values at the low metallicity end
and lower at the high end (Table~\ref{tab-means}, Figure~\ref{f2}).
The new metallicities thus steepen the slope of the RRc metallicity-period
relationship.
Additionally, the slope appears to be different in the MR and MP regimes
(again, with either the K13 or the new metallicities).
If one ignores our defined MR/MP split at [Fe/H]~=~$-$1.0, the MP
metallicity-period relationship could extend up to [Fe/H]~$\sim$~$-$0.8;
the slope change may affect only the most metal-rich RRc stars.
We do not have a physical explanation for this effect, and suggest that
more MR RRc stars be identified and studied at high spectral resolution 
in the future.

A medium-resolution spectroscopic survey by \cite{layden94} 
derived [Fe/H] metallicities for about 300 field RRab stars.
In that work, measured ``psuedoequivalent widths'' of the K-line and Balmer 
lines H$\beta$, H$\gamma$, and H$\delta$ were transformed into metallicities
using the high-resolution [Fe/H] abundances of \cite{butler75} and 
\cite{butler82}, and ultimately tied to the globular cluster metallicity scale
of \cite{zinn84}.
\cite{chadid17} analyzed 28 newly acquired RRab du~Pont echelle spectra
and reconsidered the abundance results from earlier high resolutions 
studies (\citealt{clementini95}, \citealt{fernley96}, \citealt{lambert96}, 
\citealt{nemec13}, \citealt{liu13}, \citealt{pancino15} and 
\citealt{for11b}).
These yielded an extensive set of [Fe/H] metallicities and [X/Fe] abundance 
ratios on an internally consistent system.
The various [Fe/H] values were correlated with those of 
\citeauthor{layden94}; see their \S3.3.
Mean regression lines in these correlations were used to suggest a common
transformation between the \citeauthor{layden94} metallicity scale and that
from high resolution spectroscopy.

\begin{figure}
\epsscale{1.00}
\plotone{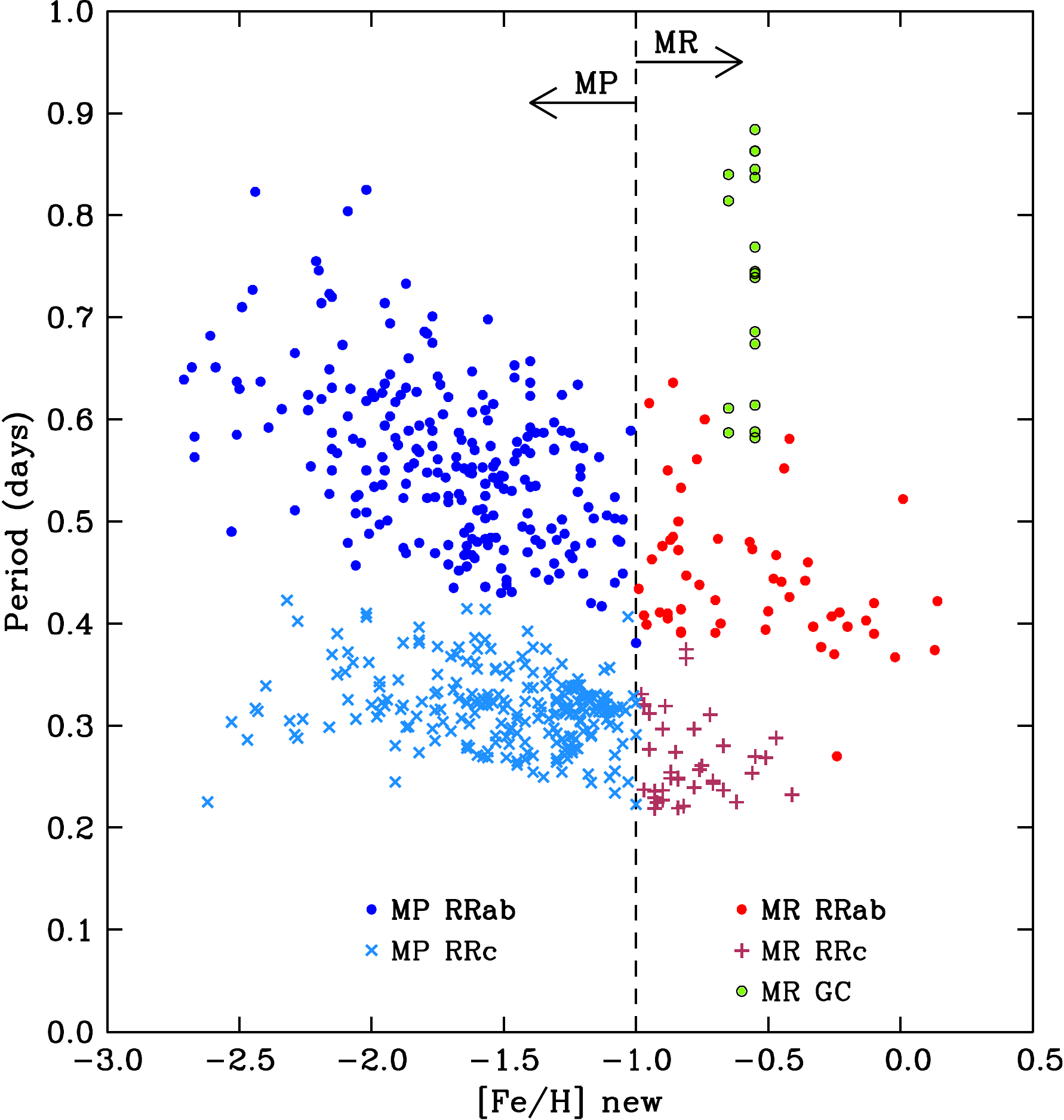}
\caption{\label{f7}
\footnotesize
   Correlation of RR~Lyrae pulsational periods with [Fe/H] 
   metallicities for our RRc stars and the RRab stars of 
   \cite{layden94,layden95b}.  
   This figure adds our program RRc stars (Table~\ref{tab-stars}) to the 
   RRab stars shown in Figure~1 of \cite{layden95b}, later adapted
   as Figure~1 of \cite{chadid17}.
   The point types and colors are given in the figure legend.
   The axis label ``[Fe/H] new'' means that the RRc metallicities are
   those of this paper, and the RRab metallicities are those of
   \cite{layden94} rescaled by \cite{chadid17}.
   The points for globular clusters (GC) are those of NGC~6388 and NGC~6441,
   two anomalous MR clusters with very extended horizontal branches
   \citep{rich97}; their [Fe/H] values are unchanged..
   The division between metal-poor (MP) and metal-rich (MR) stars at
   [Fe/H]~=~$-$1.0 is as in Figure~\ref{f6}.
   See the text for more discussion of the metallicity scales.
}
\end{figure}

The trend with metallicity for the \cite{chadid17} data can be described
well by a linear relationship,
[Fe/H]$_{Chadid}$~= 1.100[Fe/H]$_{Layden}$ + 0.055 .
We have used this formula to shift the \cite{layden94} [Fe/H] values
on to a scale that should be more consistent with the one we have derived
for the RRc stars of the present study.
In Figure~\ref{f7} we reproduce \citeauthor{layden95b}'s Figure~1 with
the recomputed RRab metallicities, and add in data for the RRc metallicities
and their ASAS periods.
Inspection of Figure~\ref{f7} suggests that RRc stars with
[Fe/H]~$\sim$~$-$2.5 have periods $P$~$\simeq$~0.35 to 0.40 days while
those with [Fe/H]~$\sim$~$-$0.2 have periods $P$~$\simeq$~0.25 to 0.30 days.
Linear regression lines for the whole data sets are:
$P_{\rm RRc} = -0.04{[\rm Fe/H]_{RRc}} + 0.25$, and 
$P_{\rm RRab} = -0.10{[\rm Fe/H]_{RRab}} + 0.39$.
Given the large star-to-star scatter in periods at all metallicities and
our adjustements to the metallicities of both RR Lyrae groups, the
slopes of these trends are similar.
Finally, we caution the reader that the RRAb and RRc sample selection 
criteria are not the same, and thus more detailed comparison of the 
period-metallicity offsets between the RRab and RRc stars is
not warranted at this time.

\acknowledgments

This work has been supported in part by NSF grant AST1616040 to C.S.  
We thank Michel Breger for helpful comments on this manuscript.

\software{IRAF (\citealt{tody93} and references therein), 
          MOOG \citep{sneden73}, 
          ATLAS \citep{kurucz11}}

%%%%%%%%%%%%%%%%%%%%%%%%%%%%%%%%%%%%%%%%%%%%%%%%%%%%%%%%%%%%%%%%%%%%%%%%%%
%    FIGURES
%%%%%%%%%%%%%%%%%%%%%%%%%%%%%%%%%%%%%%%%%%%%%%%%%%%%%%%%%%%%%%%%%%%%%%%%%%

\clearpage

\clearpage

\clearpage

\clearpage

\clearpage

\clearpage                                                   

\clearpage

%%%%%%%%%%%%%%%%%%%%%%%%%%%%%%%%%%%%%%%%%%%%%%%%%%%%%%%%%%%%%%%%%%%%%%%%%%
%   TABLES
%%%%%%%%%%%%%%%%%%%%%%%%%%%%%%%%%%%%%%%%%%%%%%%%%%%%%%%%%%%%%%%%%%%%%%%%%%

\clearpage
\begin{center}
\begin{deluxetable}{cccccccc}
\tabletypesize{\footnotesize}
\tablewidth{0pt}
\tablecaption{RRc Program Stars\label{tab-stars}}
\tablecolumns{8}
\tablehead{
\colhead{Star Name\tablenotemark{a}}             &
\colhead{$P$ (d)}                                &
\colhead{$HJD_0$ (d)}                            &
\colhead{$V_{max}$\tablenotemark{b}}             &
\colhead{$V_{amp}$\tablenotemark{c}}             &
\colhead{$S/N$\tablenotemark{d}}                 &
\colhead{[Fe/H]}                                 &
\colhead{stack \#\tablenotemark{e}}              \\
\colhead{ASAS}                                   &
\colhead{ASAS}                                   &
\colhead{ASAS}                                   &
\colhead{ASAS}                                   &
\colhead{ASAS}                                   &
\colhead{K13}                                 &
\colhead{K13}                                 &
\colhead{}         
}
\startdata
\mbox{124115-4056.9}     &     0.232417 &     1871.657 &    12.18 &     0.35 &      9.9 &  $+$0.04 &         1 \\
\mbox{012052+2143.7}     &     0.287790 &     2626.416 &    10.81 &     0.37 &      8.2 &  $-$0.04 &         1 \\ 
\mbox{095845-5927.8}     &     0.268609 &     1869.479 &    11.88 &     0.50 &     12.4 &  $-$0.10 &       no  \\
\mbox{141107-4212.2}     &     0.269753 &     1888.518 &    12.40 &     0.41 &      8.7 &  $-$0.15 &         1 \\
\mbox{195123+0835.0}     &     0.253592 &     2383.449 &    10.55 &     0.11 &     20.3 &  $-$0.16 &       no  \\
\mbox{054810-2001.4}     &     0.225147 &     1869.255 &     8.16 &     0.34 &     25.6 &  $-$0.24 &       no  \\
\mbox{210841-5509.1}     &     0.280250 &     1871.323 &    13.36 &     0.54 &      9.5 &  $-$0.31 &         1 \\
\mbox{045648+1818.3}     &     0.236865 &     2622.164 &    11.54 &     0.37 &     22.4 &  $-$0.31 &         1 \\
\mbox{185248-5113.8}     &     0.246210 &     1965.345 &    13.26 &     0.39 &     12.8 &  $-$0.36 &         1 \\
\mbox{042421+0048.8}     &     0.243670 &     1929.411 &    12.47 &     0.31 &     21.9 &  $-$0.36 &         1 \\
\mbox{203145-2158.7}     &     0.310712 &     1874.096 &    11.27 &     0.37 &     13.7 &  $-$0.38 &         2 \\
\mbox{083947+1417.4}     &     0.260931 &     2623.553 &    11.64 &     0.37 &     18.4 &  $-$0.41 &         2 \\
\mbox{071045+1059.2}     &     0.256969 &     2387.941 &    12.73 &     0.39 &     16.0 &  $-$0.43 &         2 \\
\mbox{082955-6434.6}     &     0.239448 &     1869.325 &    12.15 &     0.57 &     16.2 &  $-$0.45 &         2 \\
\mbox{224249-7430.3}     &     0.296880 &     1870.379 &    11.92 &     0.18 &     16.4 &  $-$0.46 &         2 \\
\mbox{143814-4025.6}     &     0.374705 &     1904.010 &    13.11 &     0.53 &     14.9 &  $-$0.49 &         2 \\
\mbox{192436+0631.4}     &     0.366142 &     2185.100 &    12.25 &     0.54 &      9.7 &  $-$0.49 &         2 \\
\mbox{115116-5548.3}     &     0.221350 &     1873.407 &    10.17 &     0.36 &     24.1 &  $-$0.51 &         2 \\
\mbox{182240-4242.9}     &     0.247590 &     1948.606 &    12.72 &     0.36 &     17.1 &  $-$0.53 &         2 \\
\mbox{175613-4346.3}     &     0.249139 &     1948.575 &    11.10 &     0.47 &     13.4 &  $-$0.54 &         2 \\
\mbox{175845-5516.4}     &     0.219404 &     1940.512 &    11.76 &     0.35 &     19.9 &  $-$0.54 &         3 \\
\mbox{045815-2244.5}     &     0.274048 &     1869.539 &    12.11 &     0.41 &     15.2 &  $-$0.55 &         3 \\
\mbox{155552-2148.6}     &     0.254144 &     1920.296 &    11.38 &     0.44 &     24.1 &  $-$0.57 &         3 \\
\mbox{185644-3622.6}     &     0.248500 &     1948.350 &     9.89 &     0.38 &     18.2 &  $-$0.58 &         3 \\
\mbox{121553-5157.4\_2}  &     0.319399 &     1871.920 &    12.25 &     0.43 &     15.5 &  $-$0.60 &         3 \\
\mbox{142032-1432.0}     &     0.296683 &     1903.631 &    12.76 &     0.42 &     18.9 &  $-$0.61 &         3 \\
\mbox{164410-0112.6}     &     0.227060 &     1936.644 &    13.31 &     0.44 &     20.1 &  $-$0.61 &         3 \\
\mbox{164128-1029.6}     &     0.236743 &     1938.338 &    12.59 &     0.45 &      8.4 &  $-$0.62 &         3 \\
\mbox{211058-2140.7}     &     0.224800 &     1873.283 &    10.26 &     0.13 &     10.0 &  $-$0.64 &         3 \\
\mbox{132225-2042.3}     &     0.235934 &     1886.414 &    10.72 &     0.40 &     30.5 &  $-$0.65 &         3 \\
\mbox{081012-2903.0}     &     0.229656 &     1869.397 &     9.53 &     0.28 &     35.9 &  $-$0.65 &         4 \\
\mbox{095438-6240.5}     &     0.218941 &     1869.191 &    12.27 &     0.54 &     15.8 &  $-$0.66 &         4 \\
\mbox{221135-4222.3}     &     0.311920 &     1871.329 &    13.13 &     0.50 &     10.7 &  $-$0.68 &         4 \\
\mbox{234439-0148.6}     &     0.276854 &     1869.174 &    12.23 &     0.38 &      4.9 &  $-$0.68 &         4 \\
\enddata
\tablenotetext{a}{Star names ending in ``\_N'', where N = 1 or 2, denote 
                   duplicate observations of the same ASAS target}
\tablenotetext{b}{$V_{max}$ is the brightest approximate $V$ magnitude during
                  the pulsational cycle}
\tablenotetext{c}{$V_{amp}$ is approximate $V$ magnitude change during
                  the pulsational cycle}
\tablenotetext{d}{mean $S/N$ of the values estimated near 4600~\AA\ and 5200~\AA}
\tablenotetext{e}{``no'' indicates a probable $\delta$~Scuti star, excluded from the stack mean}

\vspace*{0.1in}
(This table is available in its entirety in machine-readable form.)
 \end{deluxetable}
\end{center}

\clearpage
\begin{center}
\begin{deluxetable}{ccccccccc}
\tabletypesize{\footnotesize}
\tablewidth{0pt}
\tablecaption{Comparison with K13\label{tab-means}}
\tablecolumns{9}
\tablehead{
\colhead{stack\#}                                &
\colhead{count}                                  &
\colhead{[Fe/H]}                                 &
\colhead{$\Delta$[Fe/H]\tablenotemark{a}}        &
\colhead{[Fe/H]}                                 &
\colhead{\teff}                                  &
\colhead{\logg}                                  &
\colhead{\vmicro}                                &
\colhead{$\langle P\rangle$}                     \\
\colhead{}                                       &
\colhead{}                                       &
\colhead{K13}                                    &
\colhead{K13}                                    &
\colhead{new}                                    &
\colhead{new}                                    &
\colhead{new}                                    &
\colhead{\kmsec,new}                             &
\colhead{d}   
}
\startdata
      1 &        7 &    -0.25 &     0.40 &    -0.49 &     6950 &      2.6 &      3.2 &      0.2567 \\
      2 &       10 &    -0.47 &     0.16 &    -0.74 &     7000 &      2.4 &      3.1 &      0.2824 \\
      3 &       10 &    -0.60 &     0.11 &    -0.79 &     7150 &      2.7 &      3.3 &      0.2537 \\
      4 &       10 &    -0.70 &     0.10 &    -0.90 &     7150 &      2.3 &      3.1 &      0.2859 \\
      5 &       10 &    -0.80 &     0.10 &    -1.02 &     7100 &      2.8 &      3.0 &      0.3049 \\
      6 &       10 &    -0.87 &     0.04 &    -1.10 &     7050 &      2.4 &      3.0 &      0.2940 \\
      7 &       10 &    -0.92 &     0.07 &    -1.05 &     7150 &      2.6 &      3.3 &      0.3198 \\
      8 &       10 &    -0.98 &     0.04 &    -1.20 &     7000 &      2.6 &      3.0 &      0.3006 \\
      9 &       10 &    -1.02 &     0.04 &    -1.24 &     7000 &      2.8 &      3.0 &      0.3145 \\
     10 &       10 &    -1.05 &     0.02 &    -1.38 &     6900 &      2.5 &      3.0 &      0.3127 \\
     11 &       10 &    -1.08 &     0.02 &    -1.20 &     7000 &      2.3 &      2.2 &      0.2996 \\
     12 &       10 &    -1.11 &     0.03 &    -1.39 &     6900 &      2.4 &      3.0 &      0.3116 \\
     13 &       10 &    -1.14 &     0.01 &    -1.47 &     6850 &      2.2 &      2.5 &      0.3031 \\
     14 &       10 &    -1.18 &     0.05 &    -1.46 &     6900 &      2.2 &      2.9 &      0.3153 \\
     15 &       10 &    -1.26 &     0.07 &    -1.35 &     7100 &      2.5 &      2.8 &      0.3246 \\
     16 &       10 &    -1.31 &     0.05 &    -1.49 &     7100 &      2.4 &      2.6 &      0.3086 \\
     17 &       10 &    -1.37 &     0.06 &    -1.49 &     6950 &      2.7 &      2.4 &      0.2981 \\
     18 &       10 &    -1.43 &     0.05 &    -1.60 &     6900 &      2.6 &      2.6 &      0.3169 \\
     19 &       10 &    -1.49 &     0.05 &    -1.54 &     6900 &      2.8 &      2.8 &      0.3337 \\
     20 &       10 &    -1.54 &     0.04 &    -1.70 &     7000 &      2.7 &      3.0 &      0.3304 \\
     21 &       10 &    -1.59 &     0.05 &    -1.52 &     7100 &      2.7 &      2.2 &      0.3200 \\
     22 &       10 &    -1.68 &     0.10 &    -1.73 &     7100 &      2.6 &      2.8 &      0.3410 \\
     23 &       10 &    -1.78 &     0.10 &    -1.70 &     7100 &      3.1 &      2.5 &      0.3236 \\
     24 &       10 &    -1.90 &     0.10 &    -1.83 &     7100 &      2.9 &      2.2 &      0.3347 \\
     25 &       10 &    -2.01 &     0.10 &    -1.95 &     7100 &      2.9 &      2.4 &      0.3123 \\
     26 &       10 &    -2.15 &     0.17 &    -2.07 &     6900 &      2.9 &      1.6 &      0.3607 \\
     27 &       10 &    -2.41 &     0.35 &    -2.14 &     7000 &      2.8 &      1.8 &      0.3372 \\
     28 &        4 &    -2.69 &     0.14 &    -2.42 &     7000 &      1.9 &      2.0 &      0.3052 \\
\enddata

\tablenotetext{a}{the breadth of K13 [Fe/H] values included in a given spectrum stack}

\end{deluxetable}
\end{center}

\end{document}